\documentclass[11 pt]{article}%
\usepackage{amsmath}%
\usepackage{amsfonts}%
\usepackage{amssymb}%
\usepackage{epsf}%
\usepackage{hyperref}%

\def\singlespace {\smallskipamount=3.75pt plus1pt minus1pt
                  \medskipamount=7.5pt plus2pt minus2pt
                  \bigskipamount=15pt plus4pt minus4pt
                  \normalbaselineskip=15pt plus0pt minus0pt
                  \normallineskip=1pt
                  \normallineskiplimit=0pt
                  \jot=3.75pt
                  {\def\smallskip {\vskip\smallskipamount}}
                  {\def\medskip   {\vskip\medskipamount}}
                  {\def\bigskip   {\vskip\bigskipamount}}
                  {\setbox\strutbox=\hbox{\vrule
                    height10.5pt depth4.5pt width 0pt}}
                  \parskip 7.5pt
                  \normalbaselines}
\def\middlespace {\smallskipamount=5.625pt plus1.5pt minus1.5pt
                  \medskipamount=11.25pt plus3pt minus3pt
                  \bigskipamount=22.5pt plus6pt minus6pt
                  \normalbaselineskip=22.5pt plus0pt minus0pt
                  \normallineskip=1pt
                  \normallineskiplimit=0pt
                  \jot=5.625pt
                  {\def\smallskip {\vskip\smallskipamount}}
                  {\def\medskip   {\vskip\medskipamount}}
                  {\def\bigskip   {\vskip\bigskipamount}}
                  {\setbox\strutbox=\hbox{\vrule
                    height15.75pt depth6.75pt width 0pt}}
                  \parskip 11.25pt
                  \normalbaselines}
\def\doublespace {\smallskipamount=7.5pt plus2pt minus2pt
                  \medskipamount=15pt plus4pt minus4pt
                  \bigskipamount=30pt plus8pt minus8pt
                  \normalbaselineskip=30pt plus0pt minus0pt
                  \normallineskip=2pt
                  \normallineskiplimit=0pt
                  \jot=7.5pt
                  {\def\smallskip {\vskip\smallskipamount}}
                  {\def\medskip   {\vskip\medskipamount}}
                  {\def\bigskip   {\vskip\bigskipamount}}
                  {\setbox\strutbox=\hbox{\vrule
                    height21.0pt depth9.0pt width 0pt}}
                  \parskip 15.0pt
                  \normalbaselines}

\begin{document}
\begin{center}
{\bf {\Large Gravitational Collapse of Inhomogeneous Dust in (2+1) Dimensions}}
\bigskip

{\large Sashideep Gutti \footnote{e-mail address: sashideep@mailhost.tifr.res.in}
}
\bigskip

{\it Tata Institute of Fundamental Research,}\\
{\it Homi Bhabha Road, Mumbai 400 005, India}
\medskip

\end{center}
\bigskip
\bigskip

\begin{abstract}
\noindent We examine the gravitational collapse of spherically symmetric inhomogeneous dust in (2+1)
dimensions, with cosmological constant, for generic initial data. We obtain analytical expressions for the interior metric. We
match the solution to a vacuum exterior. We discuss the nature of the singularity formed by analyzing the outgoing radial null geodesics. We examine the formation of
trapped surfaces during the collapse.
\end{abstract}
\singlespace
\section{Introduction}
The analysis of (2+1) dimensional gravitational collapse models is useful as a toy model which may provide useful insights regarding the validity of Cosmic Censorship Hypothesis(CCH). There is no general proof for the validity of the hypothesis. There are many examples in which naked singularity is the outcome of the gravitational collapse. A good overview of the gravitational collapse scenario can be found in some of the review articles, Wald \cite{wald}, Harada \cite{harada}, Singh \cite{matsci}, Joshi \cite{joshi}. We can ask what the (2+1) dimensional Einstein's gravity has to say about CCH.

 The study is important from the point of view of quantizing general relativity. The quantum gravitational effects play an important role in the gravitational collapse, especially when the collapse results in the formation of naked singularities. A lot of work has been done based on the canonical quantization of the LeMaitre-Tolman-Bondi(LTB) spherical dust collapse models \cite{midi}. The canonical dynamics of the collapsing dust is analyzed by embedding the spherically symmetric ADM 4-metric in the LTB space time. Extending the quantization program to (2+1) dimensions is important due to the potential insights it might provide regarding the end state of the quantum gravitational collapse. This paper focuses mainly on obtaining the exact solutions of collapse of inhomogeneous spherical dust in (2+1) Einstein's theory of gravity, which can be useful for analyzing the quantum gravitational analogue.  

 The (2+1) dimensional gravity is fascinating in it's own right.  In this case there is no gravity outside matter. Matter curves space time only locally. Consequently there are no gravitational waves. The  correspondence of Einstein's theory with Newtonian gravity breaks down. Newtonian gravity cannot be obtained as a limit of Einstein's theory. In the absence of a cosmological constant a dust distribution moves without any geodesic deviation between particles \cite{gidd}. These features make the (2+1) dimensional gravity an interesting toy model.

The work on homogeneous collapse of dust in (2+1) dimensions with a cosmological constant was done by Ross and Mann \cite{ross}. They have shown that the stationary black hole solution found in \cite{bandos} arises naturally from the gravitational collapse of pressureless dust for a negative cosmological constant. It's properties were shown to be similar to the higher dimensional  Oppenheimer-Snyder case. They have shown that there is no black hole formation for the case when cosmological constant is zero, agreeing with the results obtained in \cite{gidd}. They have also shown that the collapse to a naked singularity is possible for the positive cosmological constant case provided that the initial density is sufficiently small. We extend the analysis for inhomogeneous spherically symmetric dust distribution starting from generic initial velocity distribution. We show that the introduction of inhomogeneities in the initial density profile do not alter the qualitative features of the homogeneous case. This is drastically different from the (3+1) dimensional case in which the introduction of inhomogeneities in the initial density profile alter the nature of singularity \cite{tpbarve}.

We assume spherically symmetric dust. For the case of dust we set the pressure to
zero. So the energy momentum tensor for dust is given by
\begin{equation}
T_{\mu\nu} = \rho u_\mu u_\nu,
\end{equation}
where $\rho$ is the density of the dust. We set up comoving coordinate
system in which the three velocity of dust is given by $u^{\mu}=
(1,0,0)$.
We assume the comoving spherically symmetric metric of the form
\begin{equation}
ds^{2}=-dt^{2} + e^{2b(t,r)}dr^{2}+R(t,r)^2d\phi^2
\label{3dmet}
\end{equation} 
where $t$ is the coordinate time, $r$ is the label of the comoving shell. $R({t,r)} $ represents
the physical radius of the collapsing shell. $\phi$ is the angular
coordinate.
The Einstein equations are 
\begin{equation}
G_{\mu\nu}+\lambda g_{\mu\nu}=\kappa T_{\mu\nu}
\label{eineqn}
\end{equation}
where positive and negative $\lambda$ corresponds to three dimensional de-Sitter space and anti-deSitter space respectively. 
The non zero components of Einstein tensor  are 
\begin{equation}
G_{tt}=\frac{e^{-2b}(b'R'-R''+e^{2b}\dot{b}\dot{R})}{R},
\label{g00}
\end{equation}
\begin{equation}
G_{tr}=\frac{R'\dot{b}-\dot{R'}}{R},
\label{g01}
\end{equation}
\begin{equation}
G_{rr}=\frac{-e^{2b}\ddot{R}}{R},
\label{g11}
\end{equation}
\begin{equation}
G_{\phi\phi}=-R^{2}(\dot{b}^{2}+\ddot{b}).
\label{g22}
\end{equation}
where $'$ is partial derivative w.r.t. $r$ and $\dot{}$ is partial derivative w.r.t. $t$.
We study the solutions for the  three cases ($\lambda$ is  zero, negative, positive) separately.

\section{Case $\lambda=0.$}
 The Einstein equations for $\lambda=0$ are as follows.
\begin{equation}
\frac{e^{-2b}(b'R'-R''+e^{2b}\dot{b}\dot{R})}{R}=\kappa\rho,
\label{E000}
\end{equation}
\begin{equation}
\frac{R'\dot{b}-\dot{R'}}{R}=0,
\label{E001}
\end{equation}
\begin{equation}
\frac{-e^{2b}\ddot{R}}{R}=0,
\label{E011}
\end{equation}
\begin{equation}
-R^{2}(\dot{b}^{2}+\ddot{b})=0.
\label{E022}
\end{equation}
Solving (\ref{E011}) gives
\begin{equation}
R=c_1t+c_2.
\label{0R1}
\end{equation}
where $c_1$ and $c_2$ are functions of $r$ alone. The functions should be at least $C^2$. The function $c_2$ can be taken to be $r$ making use of the scaling freedom. This implies when $t=0, R=r$. $c_1$ is interpreted as the initial velocity of the particular shell with label $r$. The counter intuitive nature of (2+1) dimensions is already apparent from the equation because if one sets the initial velocity to zero, the entire dust cloud remains at rest without collapsing. The dust cloud always moves with uniform velocity. The behavior of dust is as if gravity were absent. This curious property can be explained because of the absence of any geodesic deviation between  dust particles in (2+1) dimensions. This is related to the fact that the Riemann tensor is completely determined by the local matter distribution \cite{gidd}. Solving equation (\ref{E022}) gives 

\begin{equation}
e^b=k_1t+k_2.
\label{0b}
\end{equation}
where $k_1$ and $k_2$ are functions of $r$ alone. One has to express these two arbitrary functions of $r$ in terms of the initial density and the initial velocity $c_1$ . 
Substituting (\ref{0R1}) and (\ref{0b}) in (\ref{E001}) gives  
\begin{equation}
c_1'k_2=k_1.
\label{0constraint}
\end{equation}
The conservation equation $ T^{0\nu}_{;\nu}=0$ gives
\begin{equation}
\rho=\frac{\psi(r)}{e^{b}R},
\label{0conserve}
\end{equation} 
where $\psi(r)$ is a function of $r$. Substituting equations (\ref{0R1}) , (\ref{0b}) in (\ref{E000}) we get 
\begin{equation}
\kappa\psi(k_1t+k_2)^2=(c_1't+1)(k_1't+k_2')-(k_1t+k_2)c_1''t+k_1c_1(k_1t+k_2)^2.
\label{0dee} 
\end{equation}
Now collecting the coefficients of equal powers of $t$, the equation (\ref{0dee}) can be brought to the form
\begin{equation}
 A(r)t^2+B(r)t+C(r)=0.
\label{0general}
\end{equation}
This implies that $A=0$, $B=0$, $C=0$. 
It can be shown that each of the preceding three equations are interrelated. Each can be derived from the other two equations using the equation (\ref{0constraint}). If we write down $C$ explicitly we get
 \begin{equation}
(\kappa\psi-k_1c_1)k_2^2=k_2' .
\label{0finleqn} 
\end{equation}
From equation (\ref{0conserve}) we get 
\begin{equation}
\psi=k_2r\rho _i ,
\label{0si} 
\end{equation}
where the symbol $\rho _i$ is used throughout the paper to represent  the initial density profile, i.e $\rho(0,r)$. The function $\rho_i(r)$ is positive (in order to satisfy the weak energy condition) and is at least $C^0$. 
Substituting in (\ref{0finleqn}) and integrating we get
\begin{equation}
\frac{1}{k_2^2}=c_1^2-2\kappa\int_{0}^{r}{\rho _i(s) sds}+c,
\label{0k2} 
\end{equation}
where $c$ is a positive constant. $c$ sets a natural scale of mass to the (2+1) dimensional universe. The constant $\kappa$ has the dimension of $(mass)^{-1}$ . This feature is different from (3+1) dimensions where $\kappa$ can be made dimensionless \cite{gidd}.
 The metric (\ref{3dmet}) becomes
\begin{equation}
ds^2=-dt^2 +\frac{(c_1't+1)^2dr^2}{c_1^2-2\kappa\int_{0}^{r}{\rho _i(s) sds}+c}+(c_1t+r)^2d\phi^2.
\label{0metric} 
\end{equation}
The functions $c_1$ and $\rho_i$ should be chosen such that $c_1^2-2\kappa\int_{0}^{r}{\rho _i(s) sds}+c>0$. 
A shell becomes singular only if initial velocity  $c_1$ is negative. The time for singularity formation for a given shell is given by  $t=-r/c_1$, which is obtained by setting $R$, the physical radius of the shell to zero. The Ricci scalar $R_i$ diverges when the physical radius goes to zero.
\begin{equation}
R_i=\frac{2e^{-2b}(b'R'-R''+e^{2b}(\dot{b}\dot{R}+R(b'^2+\ddot{b})+\ddot{R}))}{R}.
\label{Ricciscalar}
\end{equation}

To answer questions about the nature of singularity obtained, we begin by looking for  trapped surfaces. The shells are trapped if the divergence of a congruence of outgoing radial null geodesics $\Theta$ is negative \cite{tpprd}. 

A congruence of outgoing radial null geodesics is considered having tangent vector $(K^t,K^r,0)$, where $K^t=dt/dk$ and $K^r=dr/dk$, where $k$ is an affine parameter along the geodesic which increases into the future. Imposing the conditions that $K^t>0$ and $K^r>0$, restricts the analysis to the case of future pointing outgoing null geodesic.
\begin{equation}
\Theta=K^i_{;i}=\frac{1}{\sqrt{-g}}\frac{\partial}{\partial{x^i}}(\sqrt{-g}K^i).
\label{Kii}
\end{equation}
For the metric (\ref{0metric}) calculation of $\Theta$ yields

\begin{equation}
\Theta=\frac{K^r}{R}(e^b\dot{R}+R').
\label{theta}
\end{equation}
Now we choose the initial velocity distribution  such that there are no shell crossing singularities. This condition can be met if $R'>0$ at least till the time when singularity is formed for that particular $r$. This implies we have to choose the function $c_1$ such that $c_1'/c_1<1/r$. 
Now evaluating $\Theta$ for the metric (\ref{0metric}) yields 
\begin{equation}
\Theta=\frac{K^r(c_1't+1)}{R}\left[\frac{c_1}{\sqrt{c_1^2-2\kappa\int_{0}^{r}{\rho _i(s) sds}+c}}+1\right].
\label{0theta}
\end{equation}
Now in the above expression $c_1't+1=R'$ which is greater than zero. We also have 
\begin{equation}
\frac{K^t}{K^r}=e^b,
\label{Ktkr}
\end{equation}
which is always positive if there are no shell crossing singularities.
$\Theta$ becomes negative only if the expression inside the square bracket becomes negative since $K^r,R,R'$ are positive. $c_1$ is negative since there would be no collapse if initial velocity is not negative. The condition for $\Theta$ to become negative is
\begin{equation}
\frac{c_1}{\sqrt{c_1^2-2\kappa\int_{0}^{r}{\rho _i(s) sds}+c}}+1<0,
\label{0thetacondn}
\end{equation}
which implies
\begin{equation}
2\kappa\int_{0}^{r}{\rho _i(s) sds}-c>0.
\label{0thetacondn2}
\end{equation}
Now we call $2\int_{0}^{r}{\rho _i(s) sds} $ the mass function. It is a monotonically increasing function. The mass function for the shell $r=0$ is zero.  Apparently, only those shells whose mass function is greater than $c/\kappa$ seem to be trapped. There is no dynamics as far as trapping is concerned since shells which are untrapped at time $t=0$ remain untrapped and vice versa. This rather unphysical property can be resolved when matching with an exterior is done.

The exterior solution for a circular symmetric (2+1) dimensional space time with a cosmological constant $\lambda$ is given by \cite{bandos}
\begin{equation}
ds^2=-(-\lambda R^2-M)dT^2+\frac{dR^2}{(-\lambda R^2-M)}+R^2d\phi^2.
\label{exterior}
\end{equation}
In the case when $\lambda=0$, $M$ should be negative in order to preserve the signature of the metric.

We set $a=-1/M$ and rescale $T$ to $\sqrt{1/a}T$, 
\begin{equation}
ds^2=-dT^2+dR^2a+R^2d\phi^2.
\label{exterior}
\end{equation}
We choose $r_0$ as the outer boundary of the collapsing dust.
Matching involves equating the first and the second fundamental forms across $r_0$  . $\phi$ is the same in the exterior and the interior metrics.
 Equating the coefficients of $d\phi^2$ on the hypersurface, we get
\begin{equation}
R_0=c_1(r_0)t+r_0.
\label{0R}
\end{equation}
Equating the remaining components
\begin{equation} 
-dt^2=-dT^2+adR^2.
\label{dteqn}
\end{equation}
This implies 
\begin{equation}
\frac{dT}{dt}=\sqrt{1+ac_1^2}.
\label{dteqn2}
\end{equation}
The unit normal to the hypersurface in the exterior metric is $ (-c_1\sqrt{a},\sqrt{a}\dot{T},0)$. In the interior metric the unit normal to the hypersurface is $(0,e^b,0)$.
Calculating the extrinsic curvature component $K^e_{\phi 
\phi}$ in the exterior metric yields
\begin{equation} 
K^e_{\phi\phi}=R\sqrt{1/a+c_1^2}.
\label{0kphie}
\end{equation}
Calculating $K^i_{\phi\phi}$ in the interior gives
\begin{equation} 
K^i_{\phi\phi}=(c_1t+r)\sqrt{c_1^2-2\kappa\int_{0}^{r_0}{\rho _i(s) sds} +c}.
\label{0kphii}
\end{equation}
Equating equations (\ref{0kphie}) and (\ref{0kphii}) we get

\begin{equation} 
a=\frac{1}{-2\kappa\int_{0}^{r_0}{\rho _i(s) sds} +c}. 
\label{0a}
\end{equation}
This implies $a$ is positive only if $c>2\kappa\int_{0}^{r_0}{\rho _i(s) sds} $. So this means that $c/\kappa$ is the upper limit on the total mass function of  the (2+1) dimensional collapsing star. 
Looking back at equation (\ref{0thetacondn2}), the above restriction on the mass implies that all the shells are always untrapped.

We now analyze the nature of the singularity. The hypersurface $R=0$ (singularity curve) is given by $t=-r/c_1$ in the $t-r$ plane.  The singularity curve is a monotonically increasing function (implied by the conditions that there are no shell crossing singularities). We identify a point on the singularity curve by  $P(t_s,r_s)$. The point $P$ on the singularity curve is said to be at least locally naked if there exists an outgoing null geodesic which terminates in the past at $P$. We restrict attention to radial outgoing null geodesics. The equation of a radial null geodesic starting at a point $P$ on the singularity curve is obtained by imposing the null condition for the metric (\ref{0metric})
\begin{equation}
\frac{dt}{dr}=\pm\frac{(c_1't+1)}{\sqrt{c_1^2-2\kappa\int_{0}^{r}{\rho _i(s) sds}+c}}.
\label{0nullg}
\end{equation}
The equation is of the form $dt/dr=f(t,r)$. Picard's theorem \cite{picard} states that if $f(t,r)$ and $\partial{f(t,r)}/\partial{t}$ are continuous functions on a closed rectangle $G$, then through each point $( 
t_s,r_s)$ in the interior of $G$ there passes a unique integral curve of the equation $dt/dr=f(t,r)$. The existence of a unique solution for the equation (\ref{0nullg}) is guaranteed by the Picard's theorem if both $f(t,r)$ (given by R.H.S of (\ref{0nullg})) and $\partial{f}/\partial{t}$ are continuous on the singularity curve and in the valid spacetime region of the $t-r$ plane. For the continuity of $f(t,r)$ and $\partial{f(t,r)}/\partial{t}$, we require $c_1$, $c_1'$ and $2\kappa\int_{0}^{r}{\rho _i(s) sds}$ be continuous. The function $c_1'$ is at least $C^1$ since $c_1$ is at least $C^2$. The function $\rho_i$ is at least $C^0$ which implies $2\kappa\int_{0}^{r}{\rho _i(s) sds}$ is at least $C^1$. We also have $c_1^2-2\kappa\int_{0}^{r}{\rho _i(s) sds}+c>0$. This implies the continuity of $f(t,r)$ and $\partial{f}/\partial{t}$ in the valid spacetime as well as on the singularity curve. The null geodesic equation (\ref{0nullg}) can be readily solved. We get
\begin{equation}
t=\pm \left[e^{\pm\int{\frac{c_1'dr}{\sqrt{c_1^2-2\kappa\int_{0}^{r}{\rho _i(s) sds}+c}}}}\right]\left[\int{\frac{e^{\mp\int{\frac{c_1'dr}{\sqrt{c_1^2-2\kappa\int_{0}^{r}{\rho _i(s) sds}+c}}}}}{\sqrt{c_1^2-2\kappa\int_{0}^{r}{\rho _i(s) sds}+c}}+b}\right],
\label{egeod}
\end{equation}
where $b$ is a constant which is fixed by choosing the initial point. A radial null geodesic equation starting from the point $P(t_s,r_s)$ can be approximated by the Taylor expansion in the neighborhood of $P$ 
\begin{equation}
t=t_s+a_0(r-r_s)+\frac{a_1(r-r_s)^2}{2!}+\frac{a_2(r-r_s)^3}{3!}+.....
\label{0taylor}
\end{equation} 
The value of $a_0$ is equal to the R.H.S of (\ref{0nullg}) (with positive sign) evaluated at the point $P(t_s,r_s)$ ,
\begin{equation}
a_0=\frac{c_1'(r_s)t_s+1}{\sqrt{c_1^2(r_s)-2\kappa\int_{0}^{r_s}{\rho _i(s) sds}+c}}.
\label{0a_0}
\end{equation}
Substituting $t_s=-r_s/c_1(r_s)$, $a_0$ can be brought to the form 
\begin{equation}
a_0=\frac{r_sc_1'(r_s)-c_1(r_s)}{c_1^2(r_s)\left[\sqrt{1 
+\frac{c-2\kappa\int_{0}^{r_s}{\rho _i(s) sds}}{c_1^2(r_s)}}\right]}.
\label{0a_0simp}
\end{equation}
Now we Taylor expand the singularity curve in the neighborhood of $P(t_s,r_s)$. We get 
\begin{equation}
t=t_s+\frac{r_sc_1'(r_s)-c_1(r_s)}{c_1^2(r_s)}(r-r_s)+...
\label{0singtay}
\end{equation}
Now the singularity will be naked if $a_0<(r_sc_1'(r_s)-c_1(r_s))/c_1^2(r_s)$. Physically this means that a null ray originating on the singularity curve at a shell with label $r_s$ is able to reach a neighboring shell $r_s+dr$ in the interval $dt$, before the shell $r_s+dr$ become singular.  Now using the equation (\ref{0singtay}) we derive that $a_0$ is less than $(r_sc_1'(r_s)-c_1(r_s))/c_1^2(r_s)$ if  
\begin{equation}
c>2\kappa\int_{0}^{r_s}{\rho _i(s) sds}.
\label{0connaked}
\end{equation}
This implies that the outgoing null rays can emerge from the singularity till the critical shell (the shell for which $c=2\kappa\int_{0}^{r}{\rho _i(s) sds}$) becomes singular. From our earlier analysis regarding the validity of the exterior spacetime, we had obtained that the condition (\ref{0connaked}) should always be valid. This implies that the singularity is always naked. The hypersurface $R=0$ is timelike. From each point on the singularity curve we obtain a unique outgoing geodesic. This is different from the (3+1) dimensional case where a family of geodesics are shown to emerge in some cases from the central singularity ($r=0$) \cite{tpbarve}. This qualitative difference is due to the fact that Picard's theorem is not applicable to the case considered in \cite{tpbarve} since $\partial{f(t,r)}/\partial{t}$ for the null geodesic equation $d 
t/dr=f(t,r)$ considered in \cite{tpbarve} is discontinuous at the central singularity. So the existence of family of solutions for the case considered in \cite{tpbarve} is not ruled out.

\section{Case $\lambda<0$}
Choose $\Lambda=-\lambda$. The Einstein equations give,
\begin{equation}
\ddot{R}+\Lambda R=0,
\label{nR}
\end{equation}
\begin{equation}
\ddot{b}+\dot{b}^2+\Lambda=0,
\label{nb}
\end{equation}
\begin{equation}
R'\dot{b}-\dot{R}'=0,
\label{nconstraint}
\end{equation}
\begin{equation}
\frac{e^{-2b}(b'R'-R''+e^{2b}(\dot{b}\dot{R}))}{R}+\Lambda=\kappa\rho.
\label{ng00}
\end{equation}
Solving equations (\ref{nR}) and (\ref{nb}) we get
\begin{equation}
R=A\cos(\sqrt{\Lambda}t)+B\sin(\sqrt{\Lambda}t),
\label{nR1}
\end{equation}
\begin{equation}
e^b=D\cos(\sqrt{\Lambda}t)+C\sin(\sqrt{\Lambda}t),
\label{nb1}
\end{equation}
where $A,B,C,D$ are functions of $r$ alone. The functions $A,B$ should be at least $C^2$. Exploiting the scaling freedom, $A$ can be set to $r$. So $R=r$ at time $t=0$. The function $B$ fixes the initial velocity profile. The functions $D,C$ have to be expressed in terms of the initial density and the initial velocity profile.
Equation (\ref{nconstraint}) implies $R'/e^b$ is a function of $r$. This implies
\begin{equation}
C=B'D.
\label{nconstrainteqn}
\end{equation}
By equation (\ref{0conserve}) we get $\rho=\psi(r)/(re^b)$. Simplification of equation (\ref{ng00}) yields

$(\Lambda(Dr+CB)-\kappa\psi)(D\cos(\sqrt{\Lambda }t)+C\sin(\sqrt{\Lambda }t))-B''\sin(\sqrt{\Lambda} t)$
\begin{equation}
+\frac{(D'\cos(\sqrt{\Lambda} t)+C'\sin(\sqrt{\Lambda }t))(\cos(\sqrt{\Lambda} t)+B'\sin(\sqrt{\Lambda }t))}{D\cos(\sqrt{\Lambda} t)+C\sin(\sqrt{\Lambda} t)}=0.
\label{nro}
\end{equation}
This equation can be brought to the form 
\begin{equation}
 X(r)\cos^2(\sqrt{\Lambda }t)+Y(r)\cos(\sqrt{\Lambda }t)\sin(\sqrt{\Lambda} t)+Z(r)\sin^2(\sqrt{\Lambda} t)=0.
\label{ngeneral}
\end{equation}
This equation is valid for all times,which implies $X=0,Y=0,Z=0$. Now $X(r)$ is
\begin{equation}
(\Lambda(Dr+CB)-\kappa\psi)D^2+D'=0.
\label{nD1}
\end{equation}
Now $\psi=\rho _irD$ where $\rho _i$ is the initial density profile of the dust.
Integrating equation (\ref{nD1}) gives 

\begin{equation}
\frac{1}{D^2}=\Lambda r^2+\Lambda B^2-2\kappa\int_{0}^{r}{\rho _i(s) sds}+c.
\label{nD2}
\end{equation}
Constant $c$ has the same interpretation as the $\Lambda=0$ case. The metric is
\begin{equation}
ds^2=-dt^2+\frac{(\cos(\sqrt{\Lambda }t)+B'\sin(\sqrt{\Lambda }t))^2dr^2}{\Lambda r^2+\Lambda B^2-2\kappa\int_{0}^{r}{\rho _i(s) sds}+c}+(r\cos(\sqrt{\Lambda }t)+B\sin(\sqrt{\Lambda }t))^2d\phi^2.
\label{nmetric}
\end{equation}
The functions $\rho_i$ and $B$ should satisfy the inequality $\Lambda r^2+\Lambda B^2-2\kappa\int_{0}^{r}{\rho _i(s) sds}+c>0$. We choose the initial velocity profile such that there can be no shell crossing singularities at least till the time singularity is reached. The condition on $B$ is $B'/B<1/r$ (assuming that the initial velocity is negative). A shell with label $r$ becomes singular when the physical radius $R$ shrinks to zero size. The shell becomes singular at time $t_s=\arctan{(-r/B)}/\sqrt{\Lambda}$. To look for trapped surfaces we analyze $\Theta$. Calculating $\Theta$ using equation (\ref{theta}) for the metric (\ref{nmetric}) yields 
\begin{equation}
\Theta=\frac{K^r(\cos(\sqrt{\Lambda }t)+B'\sin(\sqrt{\Lambda }t))}{R} \left[1+\frac{\sqrt{\Lambda}(-r\sin(\sqrt{\Lambda }t)+B\cos(\sqrt{\Lambda }t))}{\sqrt{\Lambda r^2+\Lambda B^2-2\kappa\int_{0}^{r}{\rho _i(s) sds}+c}}\right].
\label{ntheta2}
\end{equation}
For a shell with label $r$ to get trapped $\Theta<0$, which implies the expression within square bracket becomes negative. The condition is
\begin{equation}
\Lambda r^2\cos^2(\sqrt{\Lambda }t)+\Lambda B^2\sin^2(\sqrt{\Lambda }t)+2\Lambda rB\cos(\sqrt{\Lambda }t)\sin(\sqrt{\Lambda }t)-2\kappa\int_{0}^{r}{\rho _i(s) sds}+c<0.
\label{nthetacond1}
\end{equation}
 Now the sum of the first three terms on the L.H.S is equal to $\Lambda R^2$. So the above inequality becomes 
\begin{equation} 
\frac{2\kappa\int_{0}^{r}{\rho _i(s) sds}-c}{\Lambda R^2}>1.
\label{ntrap}
\end{equation}
So this implies only those shells whose mass function is greater than $c/\kappa$ can ever become trapped.  For the outer shells which can get trapped, trapping occurs when the the physical radius shrinks to a size $R<\sqrt{(2\kappa\int_{0}^{r}{\rho _i(s) sds}-c)/\Lambda}$. This is different from a (3+1) dimensional case where a black hole can be formed from any amount of mass by shrinking it to a radius less than the Scwharzschild radius. In (2+1) dimensional case if a star has the total mass function less than $c/\kappa$, there is no formation of trapped surfaces. 

To analyze the nature of the singularity formed, we use the argument similar to the case $\lambda=0$. Now the hypersurface $R=0$ is the singularity curve in the $t-r$ plane. It is given by the equation
\begin{equation}
 t=\arctan{(-r/B)}/\sqrt{\Lambda}.
\label{scur}
\end{equation}
Now let $P(t_s,r_s)$ be a point on the singularity curve. The curve can be Taylor expanded near the point $P$. We get
\begin{equation}
t=t_s+\frac{r_sB'(r_s)-B(r_s)}{\sqrt{\Lambda}(B^2(r_s)+r_s^2)}(r-r_s)+....
\label{ntaylorsingcur}
\end{equation} 
 The null condition for the metric (\ref{nmetric}) is 
\begin{equation}
\frac{dt}{dr} =\frac{\cos(\sqrt{\Lambda }t)+B'\sin(\sqrt{\Lambda }t)}{\sqrt{{\Lambda r^2+\Lambda B^2-2\kappa\int_{0}^{r}{\rho _i(s) sds}+c}}}.
\label{nnullg} 
\end{equation}
Picard's theorem guarantees a unique solution for the geodesic starting from the point $P(t_s,r_s)$. Near the point $P$, we Taylor expand the solution,
\begin{equation}
t=t_s+a_0(r-r_s)+\frac{a_1(r-r_s)^2}{2!}+....
\label{ngeodassume}
\end{equation}
 The value of $a_0$ is the R.H.S of equation (\ref{nnullg}) evaluated at $P(t_s,r_s)$ where $t_s=\arctan{(-r_s/B(r_s))}/\sqrt{\Lambda}$. We obtain,  
\begin{equation}
a_0=\frac{B'(r_s)r_s-B(r_s)}{(B^2(r_s)+r_s^2)\sqrt{\Lambda}\sqrt{\left[1+\frac{c-2\kappa\int_{0}^{r_s}{\rho _i(s) sds}}{\Lambda (B^2(r_s)+r_s^2)}\right]}}.
\label{nKr1}
\end{equation}
Comparing with equation (\ref{ntaylorsingcur}) we obtain the condition that the singularity is naked if 
\begin{equation}
a_0=\frac{B'(r_s)r_s-B(r_s)}{(B^2(r_s)+r_s^2)\sqrt{\Lambda}\sqrt{\left[1+\frac{c-2\kappa\int_{0}^{r_s}{\rho _i(s) sds}}{\Lambda (B^2(r_s)+r_s^2)}\right]}}<\frac{r_sB'(r_s)-B(r_s)}{\sqrt{\Lambda}(B^2(r_s)+r_s^2)}.
\label{ncondfornakedness}
\end{equation}
Now this is satisfied only if 
\begin{equation}
c>2\kappa\int_{0}^{r_s}{\rho _i(s) sds}.
\label{nconnaked}
\end{equation}

If we consider a star whose total mass function is greater than $c/\kappa$, we identify the critical shell $r_c$ for which $\int_{0}^{r_c}{\rho _i(s) sds}=c/\kappa$. Outgoing null rays can emerge from the singularity till the time when the critical shell becomes singular. This implies that during the collapse, the singularity will be at least locally naked till the time when the critical shell becomes singular.
The $R=0$ hypersurface is therefore timelike for the shells $r<r_c$. It becomes null for $r=r_c$ and is spacelike for $r>r_c$. 

The exterior in (2+1) dimensions is given by the BTZ metric \cite{bandos} with zero angular momentum,
\begin{equation}
ds^2=-(\Lambda R^2-M)dT^2+\frac{dR^2}{(\Lambda R^2-M)}+R^2d\phi^2.
\label{nexterior}
\end{equation}
We denote the outer boundary of the dust by the coordinate $r_0$. Matching the metric components of (\ref{nmetric}) and (\ref{nexterior}) on the hypersurface
$r_0$ gives
\begin{equation}
R_0=r_0\cos(\sqrt{\Lambda }t)+B(r_0)\sin(\sqrt{\Lambda }t),
\label{nRmat}
\end{equation}
\begin{equation}
1=(\Lambda R^2-M)\dot{T}^2-\frac{\dot{R}^2}{(\Lambda R^2-M)}.
\label{nphimat}
\end{equation}
Unit normal to the hypersurface in the interior coordinates is $(0,e^b,0)$ and in the exterior coordinates is $(-\dot{R},\dot{T},0)$.
Calculating $K_{\phi\phi}$ in the interior gives 
\begin{equation}
K^i_{\phi\phi}=e^{-b}R(\cos(\sqrt{\Lambda }t)+B'\sin(\sqrt{\Lambda }t)).
\label{nkiphi}
\end{equation}

Calculating $K_{\phi\phi}$ in the exterior gives 
\begin{equation}
K^e_{\phi\phi}=\dot{T}R(\Lambda R^2-M).
\label{nkephi} 
\end{equation}
Equating both we get
\begin{equation}
M=2\kappa\int_{0}^{r_0}{\rho _i(s) sds}-c.
\label{nM}
\end{equation}
So in the exterior static solution, $M$ can be both positive and negative. The singularity will be naked if $M$ is negative otherwise it will be a black hole \cite{ross} with event horizon at $R=\sqrt{M/\Lambda}$.

\section{Case $\lambda>0$}
Solving the Einstein equations for the metric (\ref{3dmet}) with a positive $\lambda$ we get
\begin{equation}
R=A\cosh(\sqrt{\lambda}t)+B\sinh(\sqrt{\lambda}t),
\label{pR1}
\end{equation}
\begin{equation}
e^b=D\cosh(\sqrt{\lambda}t)+C\sinh(\sqrt{\lambda}t),
\label{pb1}
\end{equation}
where $A,B,C,D$ are functions of $r$ alone. $A$ can be set to $r$. So $R=r$ at time $t=0$. The function $B$ fixes the initial velocity profile. Solution of Einstein's equations proceeds much the same way as the negative $\lambda$ case. The metric therefore is
 \begin{equation}
ds^2=-dt^2+\frac{(\cosh(\sqrt{\lambda }t)+B'\sinh(\sqrt{\lambda }t))^2dr^2}{-\lambda r^2+\lambda B^2-2\kappa\int_{0}^{r}{\rho_i(s) sds}+c}+(r\cosh(\sqrt{\lambda }t)+B\sinh(\sqrt{\lambda }t))^2d\phi^2.
\label{pmetric}
\end{equation}
There is a restriction on $B$ which comes from the fact that in the above metric, the signature has to be preserved.
\begin{equation}
 -\lambda r^2+\lambda B^2-2\kappa\int_{0}^{r}{\rho_i(s) sds}+c>0.
\label{cond1}
\end{equation}
A shell with label  $r$ becomes singular only if it's initial velocity is sufficiently high to overcome the effect of positive cosmological constant. More precisely, when  $R$  becomes zero , $B=-r/(\tanh{\sqrt{\lambda}t_s})$ where $t_s$ is the time for singularity formation. This implies that for singularity formation, it is necessary that $B<-r$ for a given shell. If $B>-r$ and $B<0$ then there will always be a rebounce at some finite $t$ and $R$ . 
The exterior is given by
\begin{equation}
ds^2=-(-\lambda R^2-M)dT^2+\frac{dR^2}{(-\lambda R^2-M)}+R^2d\phi^2.
\label{pexterior}
\end{equation}

We get the right signature only if $M$ is negative. In this case the cosmological horizon occurs at $ R_c=\sqrt{-M/\lambda}$. 
 We restrict our analysis to the case in which all shells collapse to singularity ($B<-r$). We choose the outer shell $r=r_0$ such that the physical radius $R$ of the outer shell is less than the cosmological horizon $R_c$.
The matching conditions on the hypersurface $r=r_0$ (outer boundary) is
\begin{equation}
M=2\kappa\int_{0}^{r_0}{\rho _i(s) sds}-c.
\label{pM}
\end{equation} 
Since $M$ is negative, $c>2\kappa\int_{0}^{r_0}{\rho _i(s) sds}$. The  
condition (\ref{cond1}) gives $B^2>r^2+M/\lambda$. The condition $B<-r$ which is required for the singularity formation, is more restrictive.

To analyze the nature of singularity, we calculate  $\Theta$ and evaluate the condition for trapping,
\begin{equation}
\Theta=\frac{K^r(\cosh(\sqrt{\lambda }t)+B'\sinh(\sqrt{\lambda }t))}{R}[1+\frac{\sqrt{\lambda}(r\sinh(\sqrt{\lambda }t)+B\cosh(\sqrt{\lambda }t))}{\sqrt{-\lambda r^2+\lambda B^2-2\kappa\int_{0}^{r}{\rho _i(s) sds}+c}}].
\label{ptheta2}
\end{equation}
The condition for trapping is 
\begin{equation}
\frac{c-2\kappa\int_{0}^{r}{\rho _i(s) sds}}{\lambda R^2}<1.
\label{ptrap}
\end{equation}
Since $c>2\kappa\int_{0}^{r}{\rho _i(s) sds}$ the above condition for trapping is never met.

 To see if there are outgoing null geodesics emerging from the singularity, we carry out the analysis as before. The singularity curve is given by  $ t=\rm{tanh^{-1}}{(-r/B)}/\sqrt{\lambda}$. Expanding the curve in the neighborhood of a point $P$ on the singularity curve we get
\begin{equation}
t=t_s+\frac{r_sB'(r_s)-B(r_s)}{\sqrt{\Lambda}(B^2(r_s)-r_s^2)}(r-r_s)+....
\label{ptaylorsingcur}
\end{equation} 
 The null condition for the metric (\ref{pmetric}) is 
\begin{equation}
\frac{dt}{dr} =\frac{\cosh(\sqrt{\lambda }t)+B'\sinh(\sqrt{\lambda }t)}{\sqrt{{-\lambda r^2+\lambda B^2-2\kappa\int_{0}^{r}{\rho _i(s) sds}+c}}}.
\label{pnullg}
\end{equation}
Picard's theorem guarantees a unique solution for the geodesic starting from the point $P(t_s,r_s)$. Near the point $P$, we approximate the solution by the series,

\begin{equation}
t=t_s+a_0(r-r_s)+\frac{a_1(r-r_s)^2}{2!}+....
\label{ptaylord}
\end{equation}
 The value $a_0$ is the R.H.S of equation (\ref{pnullg}) evaluated at $P(t_s,r_s)$ where $t_s=\tanh^{-1}{(-r_s/B(r_s))}/\sqrt{\lambda}$. We obtain, 
\begin{equation}
a_0=\frac{B'(r_s)r_s-B(r_s)}{(B^2(r_s)-r_s^2)\sqrt{\lambda}\sqrt{\left[1+\frac{c-2\kappa\int_{0}^{r_s}{\rho _i(s) sds}}{\lambda (B^2(r_s)-r_s^2)}\right]}}.
\label{nKr}
\end{equation}
Comparing with equation (\ref{ptaylorsingcur}) we obtain the condition that the singularity is naked if 
\begin{equation}
a_0=\frac{B'(r_s)r_s-B(r_s)}{(B^2(r_s)-r_s^2)\sqrt{\lambda}\sqrt{\left[1+\frac{c-2\kappa\int_{0}^{r_s}{\rho _i(s) sds}}{\lambda (B^2(r_s)-r_s^2)}\right]}}<\frac{r_sB'(r_s)-B(r_s)}{\sqrt{\lambda}(B^2(r_s)-r_s^2)}.
\label{pcondfornakedness}
\end{equation}  
For the above condition to be satisfied we require  
\begin{equation}
\frac{c-2\kappa\int_{0}^{r_s}{\rho _i(s) sds}}{\lambda (B^2(r_s)-r_s^2)}>0.
\label{pnullcondn}
\end{equation}
For singularity formation, we derived earlier that $B<-r$. So $B^2>r^2$. So the above inequality is met if
\begin{equation}
c>2\kappa\int_{0}^{r_s}{\rho _i(s) sds}.
\label{pconnaked}
\end{equation}
Since this condition is always met in the collapse scenario under consideration,  outgoing null rays can emerge from the singularity. The hypersurface $R=0$ is timelike. The singularity is therefore naked. 

It is interesting to note that for $3+1 $  dimensional collapse  with positive $\lambda$, there are usually two horizons formed. One is the black hole event horizon and the other is the cosmological  horizon \cite{markovic}.

\section{Conclusions}

\indent Dust in (2+1) dimensions can undergo a variety of collapse scenarios. The curvature singularity formed due to the collapse of dust can be both naked or covered, depending on the sign of the cosmological constant and the initial density profile of the dust. In the absence of cosmological constant, collapse to a singularity is possible only if the initial velocity of dust is negative. The matching across the dust edge between the interior solution and the flat exterior is possible only if the total mass function of the collapsing dust ($2\int^{r_0}_0{\rho_i(s) sds}$,  where $r_0$ is the outer boundary) has an upper bound. The analysis of $\Theta$ (the divergence of outgoing null geodesics) rules out the formation of trapped surfaces during collapse. The singularity formed is naked since outgoing null rays are found to emerge from the singularity.  

The collapse with a negative cosmological constant has several interesting features. In this scenario, collapse to a singularity is always possible. The analysis of $\Theta$ indicates the existence of a lower bound on the mass function of the shells that can become trapped during the collapse. The exterior static solution (BTZ blackhole with zero angular momentum) can be matched to the interior metric. The matching across the dust edge gives the relationship between the  parameter $M$ (analogous to the Schwarzschild mass) in the BTZ exterior, and the mass function of the outer shell. From the analysis of outgoing null geodesics emerging from the singularity, it is concluded that the singularity is either naked or covered depending on the mass function of the collapsing shell. The singularity is covered only if the mass function of the shell is greater than a critical value. 

 Collapse to a singularity occurs in the presence of a positive cosmological constant only if the initial velocity of the dust shell is high enough to overcome the repulsive effect of the positive cosmological constant. The collapse results in the formation of naked singularity.
\section{Acknowledgment}
I would like to thank T.P.Singh, Rakesh Tibrewala, Suresh Nampuri, Arun Madhav T Menon and Ashutosh Mahajan, for useful discussions.

\end{document}